\documentclass[preprint]{elsarticle}
\usepackage[utf8]{inputenc}

\usepackage{amssymb}
\usepackage{amsthm}
\usepackage{amsmath}
\usepackage{dsfont}
\usepackage[usenames,dvipsnames]{xcolor}
\usepackage{booktabs}
\usepackage{graphicx}
\usepackage{appendix}
\usepackage{bm}
\usepackage{soul}

\newcommand{\eb}{\bm{e}}

\newcommand{\ub}{\bm{ u}}

\newcommand{\wb}{\bm{ w}}

\newcommand{\p}{\partial}

\newcommand{\alphab}{\boldsymbol{\alpha}}
\newcommand{\lambdab}{\bm{\lambda}}
\newcommand{\psib}{\boldsymbol{\psi}}

\newcommand{\ba}{\begin{array}}
\newcommand{\ea}{\end{array}}
\newcommand{\be}{\begin{equation}}
\newcommand{\ee}{\end{equation}}
\newcommand{\bd}{\begin{displaymath}}
\newcommand{\ed}{\end{displaymath}}

\newcommand{\R}{\mathds{R}}
\newcommand{\C}{\mathds{C}}

\title{A time-parallel multiple-shooting method for large-scale quantum optimal control\tnoteref{t1}}
\tnotetext[t1]{This work was performed under the auspices of the U.S.~Department of Energy by Lawrence Livermore National Laboratory under Contract DE-AC52-07NA27344. LLNL-JRNL-866412.}

\author[a]{N.~Anders Petersson\corref{cor1}}
\ead{petersson1@llnl.gov}

\author[a]{Stefanie G{\"u}nther}
\ead{guenther5@llnl.gov}

\author[a]{Seung Whan Chung}
\ead{chung28@llnl.gov}

\affiliation[a]{organization={Center for Applied Scientific Computing, Lawrence Livermore National Laboratory},
addressline={L-561, 7000 East Ave.},
city={Livermore}, 
postcode={CA 94550},
country={USA}}

\cortext[cor1]{Corresponding author}

\begin{document}

\begin{abstract}
  Quantum optimal control plays a crucial role in quantum computing by
  providing the interface between compiler and hardware. Solving the optimal control problem is particularly challenging for multi-qubit gates, due to the exponential growth in computational complexity with the system's dimensionality and the deterioration of optimization convergence. To ameliorate the computational complexity of time-integration, this paper introduces a  multiple-shooting approach in which the time domain is divided into multiple windows and the intermediate states at window boundaries are treated as additional optimization variables. This enables parallel computation of state evolution across time-windows, significantly accelerating objective function and gradient evaluations. Since the initial state matrix in each window is only guaranteed to be unitary upon convergence of the optimization algorithm, the conventional gate trace infidelity is replaced by a generalized infidelity that is convex for non-unitary state matrices. Continuity of the state across window boundaries is enforced by equality constraints. A quadratic penalty optimization method is used to solve the constrained optimal control problem, and an efficient adjoint technique is employed to calculate the gradients in each iteration. We demonstrate the effectiveness of the proposed method through numerical experiments on quantum Fourier transform gates in systems with 2, 3, and 4 qubits, noting a speedup of 80x for evaluating the gradient in the 4-qubit case, highlighting the method's potential for optimizing control pulses in multi-qubit quantum systems.
\end{abstract}


\maketitle

\section{Introduction}

Quantum optimal control serves a prominent role in quantum computing applications, where control pulses are used to steer the quantum system to realize gate transformations, serving as the interface between compiler and hardware~\cite{Debnath-16}.
Leveraging classical computing, quantum optimal control employs numerical optimization to find control pulses that minimize the errors in fundamental operation~\cite{glaser2015training, koch2022quantum}. When applied to larger multi-qubit operations, it holds the potential to realize high-fidelity multi-qubit gates with minimal errors and shorter durations compared to standard circuit compilation~\cite{spiteri2018quantum, smith2022programming}. However, the quantum optimal control problem is computationally challenging for three main reasons. First, the numerical methods must be chosen carefully to properly reflect the physical properties of the quantum system, such as conservation of total probability and preservation of unitary evolution. Secondly, the computational complexity grows quickly with the dimension of the underlying Hilbert space, because of the exponential scaling in the number of coupled sub-systems (qubits). The third computational challenge is that the convergence of the optimization algorithm can deteriorate when more than two coupled systems are being controlled, for example, to realize multi-qubit unitary transformations.

Several different approaches have been proposed for numerically solving the open-loop quantum optimal control problem. In the gradient-free CRAB algorithm~\cite{Caneva-2011}, the number of control parameters is kept small by parameterizing the control pulses using a small number of basis functions (Fourier, Lagrange, sinc, etc), resulting in a multivariate optimization problem that can be solved using a direct search method. 
For more general control pulses, where larger numbers of control parameters are needed, quasi-Newton optimization methods usually converge faster. Here the gradient with respect to the control parameters can be calculated by solving an adjoint state equation backwards in time. The GRAPE algorithm~\cite{Khaneja-2005,Leung-2017} first discretizes the control pulses to be constant within each time step, and then optimizes the corresponding control amplitudes. Another popular approach is Krotov's method~\cite{Morzhin2018KrotovMF, goerz2019krotov}, in which the optimal control algorithm is first formulated in terms of continuous control functions and then discretized in time. Here the control energy is penalized to regularize the optimization problem, guaranteeing that the objective (cost) function decreases monotonically during the optimization iteration. A combination of these approaches, where the continuous control functions are first expanded in terms of B-spline wavelets, leads to an optimal control problem for the B-spline coefficients. This allows the number of control parameters to be significantly smaller than the number of time-steps, and also allows the gradient of the objective function to be calculated exactly, leading to an efficient optimal control algorithm~\cite{gunther2021quandary, PetGar-22} that is implemented in the High-Performance-Computing (HPC) centric Quandary code~\cite{quandaryGithub}.

All of the above optimal control methods belong to the class of reduced-space optimization methods, in which the vector of control parameters constitutes the sole design (optimization) variables. In a reduced-space method, the state equation (Schr\"odinger's equation for a closed quantum system) is satisfied for each control vector, and the gradient can be calculated by solving the adjoint state equation. This implies that in each iteration of the optimization algorithm, the state evolution is feasible, i.e., satisfies the constraints imposed by the state equation. The reduced-space approach works well for single and two-qubit gates and moderate gate durations, but is less suited for controlling larger quantum systems.
Two problems occur when the dimension of the Hilbert space gets larger. First, the computational costs for solving the state and adjoint equations increase rapidly, as longer pulse durations are required when controlling multi-qubit systems, requiring more and more time steps to resolve the underlying dynamics in each optimization iteration (cf., for example, the test cases in Table \ref{tab:testcases} (\S~\ref{sec_num-exp})). Secondly, the optimizer converges more slowly, i.e., more iterations are required to solve the optimal control problem. This paper tackles the former by introducing computational parallelism along the time domain.

As alternative to the reduced-space method, a direct collocation method~\cite{Trowbridge-23} can be used for solving the quantum optimal control problem. 
In this method, the state equation is first discretized in time by a collocation scheme and the states at the beginning of each time step are treated as additional optimization variables, hence belonging to the class of full-space optimization methods~\cite{borzi2011computational}. The continuity of the state between time steps is then imposed through equality constraints. Because the constraints are only guaranteed to be satisfied upon convergence (up to a prescribed tolerance), the state equation is in general not satisfied during the optimization iteration.
The advantage of a collocation method is that the time-stepping decouples, and can be performed locally in time. However, the number of optimization variables can become very large, which increases the memory requirements and may make the optimization problem hard to solve. 

\begin{figure}[htp]
\centering
    \includegraphics[width=0.65\textwidth]{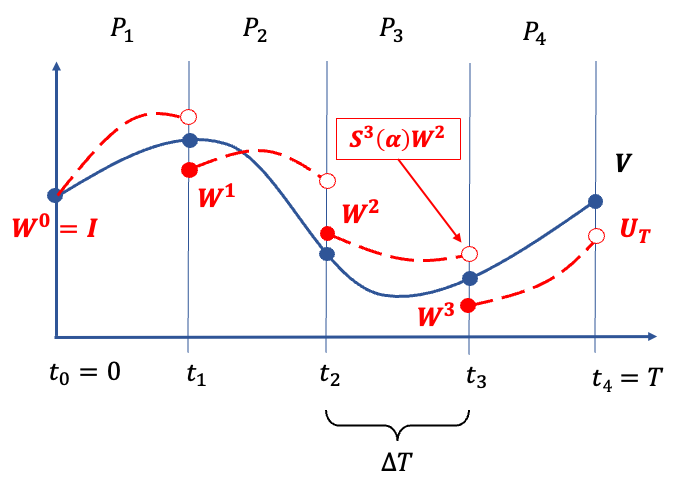}
\caption{Notation for the multiple-shooting approach. In this case, there are $M=4$ time windows and $3$ intermediate initial conditions: $W^{1}$, $W^{2}$, $W^{3}$. The blue line shows the state evolution for the converged (continuous) solution and the dashed red lines exemplify the (discontinuous) state evolution corresponding to an intermediate optimization iteration. The evaluation of the objective and its gradient can performed concurrently for each time window, in this case on four computational processes: $P_1,\ldots,P_4$. 
}
\label{fig:time-int}
\end{figure}

In this paper, we develop an intermediate optimization strategy for quantum optimal control based on multiple-shooting~\cite{ChuFre-22, JanMey-24, PucBaiBeg-23}. Here the time domain is split into several segments (windows) and optimization variables are introduced for representing the state at the beginning of each time window, see Figure~\ref{fig:time-int}.
Continuity of the state evolution across window boundaries is enforced through equality constraints on the state. 
Compared to the conventional reduced-space approach, one significant advantage of the multiple-shooting strategy is that the state (and adjoint) equation can be solved independently and concurrently in each time window. This enables significantly faster evaluations of the objective function and its gradient, through parallel-in-time processing using classical HPC platforms. The multiple-shooting strategy only adds optimization variables at the beginning of each time window, thus limiting the memory footprint and introducing a smaller number of equality constraints, compared to the direct collocation method.
Time-parallel implementations of the multiple-shooting method have recently been reported for optimal control of incompressible fluid flow~\cite{JanMey-24}, as well as for models of turbulent flow~\cite{ChuFre-22}. Algorithmic speed-ups, i.e., reduction in the number of optimization iterations, have been reported for multiple-shooting methods applied to large scale optimal control problems governed by parabolic PDEs~\cite{Fang-22}. 

As an initial evaluation of the multiple-shooting approach applied to the quantum optimal control problem, we here use the quadratic penalty method~\cite{nocedal2006numerical} to reformulate the equality constraints across window boundaries into penalty terms that are added to the objective function. 
The overall computational cost of the proposed multiple-shooting algorithm essentially follows as the cost of performing one iteration, which is dominated by the time-stepping in each window for computing the objective and its gradient, times the number of optimization iterations for reaching an optimal solution. The basic strategy of the multiple-shooting approach is to choose the number of windows to minimize the overall cost for solving the optimization problem. For example, when the number of time windows is large, the time-stepping cost per iteration can be small due to significant parallel speed-up, but the number of optimization variables and equality constraints will be large, potentially resulting in a large number of optimization iterations for finding an optimal solution. If only one time window is used, the multiple-shooting method reverts to a conventional reduced-space method. On the other extreme, if the number of time windows equals the number of time steps, the multiple-shooting method resembles a direct collocation method. Thus, the multiple-shooting method bridges the gap between the extremes by allowing the number of time windows to be chosen freely.

In this paper we focus on unitary gate transformations in a closed quantum system. The governing equations are outlined in \S\ref{sec_governing-eqn}, where we also introduce a control parameterization based on carrier waves with smoothly varying envelopes represented by B-splines. 
Our formulation of the multiple-shooting method is introduced in \S\ref{sec_multiple-shooting}.
While it is natural to enforce the state evolution to be unitary in a reduced-space optimization formalism, it is less natural to do so in a multiple-shooting setting. This is because the state matrices that represent the intermediate initial conditions will in general not be unitary during the optimization iteration. One option would be to project these state matrices onto the unitary manifold, however, at the cost of significantly increasing the computational complexity. Here we choose another approach and enlarge the state space to include complex-valued matrices. As a consequence, the conventional definition of the gate infidelity must be modified, because it can become unbounded for non-unitary state matrices. To mitigate this problem, we propose a small modification, resulting in a generalized infidelity that is non-negative and convex for non-unitary state matrices.
In \S~\ref{sec_adjoint-grad}, we derive the adjoint-based formulations for calculating the gradient of the objective function with respect to the controls and the intermediate initial conditions. In particular, we show that the adjoint state, which is used to calculate the gradient with respect to the control vector, can also be used to calculate the gradient with respect to the state at the beginning of each time window. As a result, all components of the gradient can be evaluated by solving one state equation and one adjoint state equation, concurrently in each window, at a cost that essentially is independent of the number of optimization variables. 

A challenge with the multiple-shooting method is to decide when to terminate the optimization iteration. 
While it would be desirable to evaluate the so called roll-out infidelity, where continuity of the state is enforced across time window boundaries, 
its calculation can {\it not} be parallelized across the time domain. Instead, in \S~\ref{sec_opt-stopping}, we derive an estimate for the roll-out infidelity that is based on the violations of the equality constraints across window boundaries, which can be evaluated locally in time.
\S\ref{sec_time-parallel} presents the time-parallelization strategy for concurrently evaluating the objective and its gradient in each time window on multiple computational processes. The resulting time-parallel algorithm has been implemented in the Quandary software~\cite{quandaryGithub} and is used in \S\ref{sec_num-exp} to evaluate parallel scaling performance of the multiple-shooting approach, first for the time-parallel calculation of the objective function and its gradient, followed by evaluating the performance of the overall optimal control algorithm based on the quadratic penalty method. Here we consider Quantum Fourier Transform (QFT) gates in quantum systems with 2, 3, and 4 coupled qubits. Conclusions and final remarks are given in \S\ref{sec_conclusions}.

\section{Governing equations}\label{sec_governing-eqn}

Let the evolution of the state of a closed quantum system be modeled as a time-dependent unitary transformation, $\psib(t) = U(t)\psib(0)$. Here, $\psib \in \mathbb{C}^n$ is the state vector and $n\geq 2$ is the dimension of the Hilbert space. The state matrix, $U\in \mathbb{C}^{n\times n}$ satisfies Schr\"odinger's equation in matrix form (scaled such that $\hbar=1$),
\begin{align}\label{eq_schro}
\dot{U}(t) = -i H(t) U(t),\quad 0\leq t\leq T,\quad U(0) = I_{n\times n}.
\end{align}
Here, $H(t) = H_s + H_c(t)$ is the Hamiltonian matrix, decomposed into a time-independent system part, $H_s$, and a time-dependent control part, $H_c(t)$. The Hamiltonian matrix is Hermitian for all times and, as a result, the state matrix $U(t)$ is unitary for all times. 
As an example, we may model $q$ superconducting qubits with dipole-dipole coupling strength $J_{jk}$, using a system Hamiltonian in a frame rotating with angular frequency $\omega^{rot}$, given by 
\begin{align}\label{eq_sys-ham}
    H_{s} = \sum_{j=1}^q \left\{\left(\omega_j - \omega^{rot}\right) A_j^\dagger A_j + 
    \sum_{k>j}^q J_{jk} \left(A_k^\dagger A_{j} + A_k A_{j}^\dagger\right)\right\}.
\end{align}
Here, $A_j\in \mathbb{C}^{n\times n}$ is the lowering matrix for sub-system $j$, and, in this case, $n=2^q$.
In the following we assume that the time-dependence in the control Hamiltonian is parameterized by a control vector $\alphab\in \mathbb{R}^d$, and is of the general form
\begin{align}\label{eq_ctrl-ham}
H_c(t;\alphab) &= \sum_{j=1}^{q} d_j(t;\alphab) A_{j} + d_j^*(t;\alphab) A_{j}^\dagger,
\end{align}
where $d_j(t;\alphab)$ is the complex-valued scalar control function and $d^*$ denotes its complex conjugate.  

Many control parameterizations have been proposed in the literature, e.g., Slepian sequences~\cite{Lucarelli-2018}, cubic splines~\cite{Ewing-1990}, Gaussian pulse cascades~\cite{Emsley-1989}, and Fourier expansions~\cite{Zax-1988}.
This paper employs an alternative parameterization using envelope functions with carrier waves~\cite{PetGar-22, gunther2021quandary}. We note that the idea of using carrier waves has been used extensively for controlling molecular systems with lasers~\cite{ShiRab-92, CombarizaEtAl-91, Brumer-92}. In this approach, the control function for the $j^{th}$ sub-system is defined by 
\begin{align}\label{eq_dj-ctrl}
    d_j(t;\alphab) = \sum_{f=1}^{N_f(j)}  \widetilde{d}_{jf}(t;\alphab) e^{it\Omega_j^f},
\end{align}
where the scalar complex-valued function $\widetilde{d}_{jf}(t)$ determines the amplitude and phase of the envelope, corresponding to carrier wave frequency $\Omega_j^f$.
The carrier wave approach relies on the observation that transitions between the energy levels in a quantum system are triggered by resonance, at frequencies that can be determined from the system Hamiltonian. 
This approach leads to narrow-band control functions, where the frequency content is concentrated near the resonant frequencies of the system. These frequencies can be identified by first diagonalizing the system Hamiltonian, $\widetilde{H}_s = X^\dagger H_s X$, and then applying the same transformation to the operators in the control Hamiltonian. Using an asymptotic expansion, it can be shown that the resonance frequencies in system $j$ are equal to certain eigenvalue differences in $\widetilde{H}_s$, with a growth rate that is determined by the elements in $X^\dagger A_j X$, see \cite{PetGar-22}, resulting in $N_f(j)$ carrier frequencies.

\begin{figure}[thp]
\centering
    \includegraphics[width=0.65\textwidth]{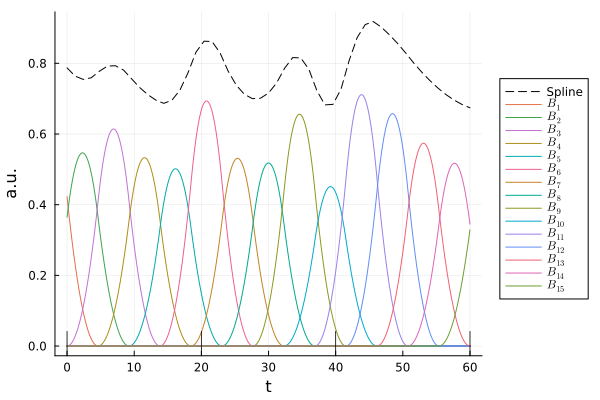}
\caption{A control envelope function (dashed-black) and its decomposition into equally spaced (cardinal) quadratic B-spline basis functions.}\label{fig:bspl-decomp}
\end{figure}

We choose to parameterize the scalar envelop functions $\widetilde{d}_{jf}(t;\alphab)$ in \eqref{eq_dj-ctrl} in terms of a set of basis functions $\{B_s(t)\}_{s=1,\dots d_1}$,
\begin{align}
    \widetilde{d}_{jf}(t;\alphab) = \sum_{s=1}^{d_1} \left( \alpha^{(1)}_{jfs} + i\alpha^{(2)}_{jfs} \right) B_s(t), \quad
\end{align}
where $\alpha_{jfs}^{(1)}$ and $\alpha_{jfs}^{(2)}$ are the real-valued control parameters in the control vector $\alphab\in\mathbb{R}^d$; the total number of elements in the control vector equals $d=2 d_1 \sum_{j=1}^q N_f(j)$. 
The are several options for choosing the basis functions in the above formula. For example, the basis functions in a Fourier or Chebyshev expansion are mutually orthogonal, and the Fourier basis has compact support in frequency space. However, using either of these bases makes it difficult to construct control functions that begin and end with zero amplitude, which may be desirable in an experimental implementation. Furthermore, the global support in time implies that each coefficient in the Fourier or Chebychev expansion modifies the control function for all times, which may hamper the convergence of the optimization algorithm. An interesting alternative is provided by semi-orthogonal B-spline wavelets, which can be optimally localized in both time and frequency~\cite{Unser97}. Here, each basis function is a smooth polynomial function of time, where the smoothness depends on the order of the spline. The B-spline basis functions are inexpensive to evaluate and have local support in time, making it easy to construct control functions that begin and end at zero. In this work we use quadratic B-splines, see Figure~\ref{fig:bspl-decomp} for an example.

A fundamental task in quantum computing is to find a control vector such that the state matrix matches a given unitary target transformation (gate) ${\cal V}\in\mathbb{C}^{n\times n}$, at a given time $T>0$ (the gate duration). Because global phase angles are inconsequential in quantum physics, the state matrix is considered to match the target if the trace infidelity,
\begin{equation}\label{eq_trace-infid}
    {\cal G}_{\cal V}\left(U(T)\right) := 1 - \frac{1}{n^2} |\langle {\cal V}, U(T)\rangle_F|^2, \quad \text{where} \quad U(T)^\dagger U(T) = {\cal V}^\dagger {\cal V} = I,
\end{equation}
vanishes.
Here, $\langle {\cal V}, U \rangle_F= \mbox{tr}({\cal V}^\dagger U)$ denotes the Frobenius matrix scalar product, with matrix norm $\| U \|_F = \langle U, U \rangle_F^{1/2}$. The trace infidelity is zero when $U(T) = \exp(i\theta) {\cal V}$, for some phase angle $\theta\in\mathbb{R}$.

The quantum optimal control problem in a reduced-space optimization method can be compactly stated as:
\begin{align}
    \min_{\alphab} {\cal G}_{\cal V}\left(U(T)\right) \quad \text{s.t.} \quad U(t) \text{ satisfies \eqref{eq_schro}.}
\end{align}
Note that the state matrix depends implicitly on the control vector, $U(t) = U(t;\alphab)$, because the Hamiltonian in \eqref{eq_schro} depends on $\alphab$.

\section{Multiple-shooting for quantum optimal control}\label{sec_multiple-shooting}

This paper introduces a quantum optimal control formulation that allows for a decomposition of the time domain into multiple time-windows, for which the state equation \eqref{eq_schro} can be solved concurrently. To this end, we split the time domain into $M > 1$ windows: $t\in[t_{m-1},t_m]$, for $m\in\{1, \dots,M\}$, with $0=t_0 < t_1 < \cdots < t_{M} = T$, see Figure~\ref{fig:time-int}. For simplicity, we assume that the durations of all windows are equal, $t_m - t_{m-1} = \Delta_T$, with $\Delta_T = T/M$. Let the initial condition in the $m$-th window be denoted by $W^{m-1}\in\mathbb{C}^{n\times n}$, for $m=1,\ldots,M$, where $W^{0} = I$ follows from the initial condition in \eqref{eq_schro} (see Figure~\ref{fig:time-int}). The initial conditions in the remaining windows are treated as additional optimization variables. 

Let the {state matrix} in window $m$ be $U^m(t)$. It satisfies Schr\"odinger's equation,
\begin{align}
    \dot{U}^m(t) + iH(t;\alpha) U^m(t) &= 0,\quad t_{m-1} < t \leq t_m, \label{eq_sch-op}\\
    U^m(t_{m-1}) &= W^{m-1},\quad W^0=I, \label{eq_sch-ic}
\end{align}
for $m=1,\ldots,M$. We will also make frequent use of the {solution operator} $S^m(t)$, which satisfies Schr\"odinger's equation \eqref{eq_sch-op} subject to the initial condition $S^{m}(t_{m-1})=I$. As a result, 
\begin{align}
    U^{m}(t) = S^{m}(t)W^{m-1},\quad m\in\{1,\dots,M\}.
\end{align} 
Note that the solution operator is unitary and depends implicitly on the control vector, i.e., $S^m(t) = S^m(t;\alphab)$. 
As a result, $U^m(t)$ is unitary if, and only if, the initial condition matrix $W^{m-1}$ is unitary. 

Splitting the time domain into multiple time windows introduces discontinuities in the state evolution across time-window boundaries.
In the multiple-shooting formalism, continuity of the state evolution is enforced by solving the following constrained optimal control problem for the control vector $\alphab$ and the intermediate initial conditions $\{W^1,\cdots,W^{m-1} \}$:
\begin{gather} 
\min_{\alphab, W^{1},\ldots,W^{M-1}} {\cal J}_{\cal V}\left(U^M(t_M)\right), \label{eq_ms_optimprob}\\
\text{s.t.} \quad
\mbox{$U^m(t)$ satisfies \eqref{eq_sch-op}-\eqref{eq_sch-ic} for $m=1,\ldots M$,} \label{eq_con1} \\
\qquad \qquad \mbox{$U^m(t_m) - W^m = \boldsymbol{0}_{n\times n}$, for $m=1,\dots, M-1$.} \label{eq_con0} 
\end{gather}
As before, $U^m(t) = U^m(t;\alphab)$. 

Because the equality constraints in \eqref{eq_con0} may not be satisfied during the optimization iteration, neither $W^{M-1}$, nor $U^M(t_M)$, are guaranteed to be unitary. As a result the trace overlap, $|\langle {\cal V}, U^M(t_M)\rangle_F|/n$, is no longer bounded by unity and the standard trace infidelity ${\cal G}_{\cal V}(U^M(t_M))$ in \eqref{eq_trace-infid} can become arbitrarily large and negative for non-unitary $U^M(t_M)$. As a mitigation, we define the generalized infidelity,
\begin{align}\label{eq_mod-infid}
{\cal J}_{\cal V}(U) &:= \frac{1}{n} \left\| U \right\|^2_F - \frac{1}{n^2}\left| \left\langle {\cal V}, U \right\rangle_F \right|^2,\quad {\cal V}^\dagger {\cal V} = I,\quad U\in\mathbb{C}^{n\times n},
\end{align}
to serve as a replacement of ${\cal G}_{\cal V}(U)$.
Note that if all of the equality constraints in \eqref{eq_con1}-\eqref{eq_con0} are satisfied, the state matrix at the final time is unitary because $U^M(t_M)= S^M(t_M) \cdots S^1(t_1) I$. When $U^M(t_M)$ is unitary, $\| U^M(t_M) \|_F^2 = n$, and the generalized infidelity simplifies to the standard trace infidelity. For general $U^M\in\mathbb{C}^{n\times n}$, Cauchy-Schwartz inequality gives $\left| \left\langle {\cal V}, U^M \right\rangle_F \right|^2 \leq \| {\cal V} \|_F^2 \| U^M \|_F^2 = n \| U^M \|_F^2$, because ${\cal V}$ is unitary, such that ${\cal J}_{\cal V}(U^M)\geq 0$. Furthermore, ${\cal J}_{\cal V}(U^M) = 0$, if and only if $U^M = \beta {\cal V}$, for some $\beta\in\mathbb{C}$. 
Additionally, the generalized infidelity ${\cal J}_{\cal V}(U)$ is a convex (but not strongly convex) function of $U\in\mathbb{C}^{n\times n}$. To prove this, consider the vectorized representations $\vec{\cal V} := \mbox{vec}({\cal V})\in C^{n^2}$ and $\vec{U}:=\mbox{vec}(U) \in C^{n^2}$, stacking the columns of each matrix into (tall) column vectors. 
Because $\langle{\cal V}, U \rangle_F = \langle \vec{\cal V}, \vec{U} \rangle$\footnote{The scalar product for vectors $\vec{u},\ \vec{v}\in\mathbb{C}^p$ is defined by $\langle\vec{u},\vec{v}\rangle = \vec{u}^\dagger\vec{v} = \sum_{k=1}^{p} u_k^* v_k$.}, the generalized infidelity \eqref{eq_mod-infid}
can be written in quadratic form,
\begin{multline}\label{eq_gen-infid}
    {\cal J}_{\cal V}(U) = \frac{1}{n}\|U\|_F^2 - \frac{1}{n^2} \langle U, {\cal V} \rangle_F \langle {\cal V}, U\rangle_F =
    \frac{1}{n} \langle \vec{U}, \vec{U}\rangle - \frac{1}{n^2}\langle \vec{U}, \vec{{\cal V}} \rangle \langle \vec{\cal V},\vec{U} \rangle \\ 
    = \frac{1}{n}\langle \vec{U}, Q_v \vec{U} \rangle,\quad 
    \text{where}\quad Q_v = I - \frac{1}{n} \vec{\cal V} \vec{\cal V}^\dagger \in\mathbb{C}^{n^2 \times n^2}.
\end{multline}
We remark that the outer product $\vec{\cal V}\vec{\cal V}^\dagger$ is a Hermitian matrix. Therefore, the matrix $Q_v$ is also Hermitian and $\langle \vec{U}, Q_v\vec{U}\rangle$ is real for all $\vec{U}\in\mathbb{C}^{n^2}$. Furthermore, the matrix $Q_v$ is a rank-1 modification of the identity matrix. As a result, all of the eigenvalues of $Q_v$, except one, are equal to unity. The remaining eigenvalue equals zero and has $\vec{\cal V}$ as eigenvector, which is easily verified by noting that $\vec{\cal V}^\dagger \vec{\cal V} = n$. Thus, $Q_v$ is positive semi-definite, which proves that ${\cal J}_{\cal V}(U)$ is a convex function of $U\in\mathbb{C}^{n\times n}$. Due to the single zero eigenvalue of $Q_v$, it is not strongly convex.

While there are many options for solving equality constrained optimization problems, see e.g. \cite{nocedal2006numerical}, we here employ the quadratic penalty method. To this end, we solve
the multiple-shooting quantum optimal control problem \eqref{eq_ms_optimprob}-\eqref{eq_con0} by minimizing the penalty objective function
\begin{gather} 
     {\cal P}(\alphab, {W}^{1},\ldots, {W}^{M-1}) = {\cal J}_{\cal V}\left(U^{M}(t_M)\right) + \frac{\mu}{2} \sum_{m=1}^{M-1} \| U^{m}(t_m) - W^{m} \|_F^2, \label{eq_penalty-fcn}
\end{gather}
where $U^m(t)$ satisfies Schr\"odinger's equation in $(t_{m-1}, t_m]$ with initial condition $U^m(t_{m-1})=W^{m-1}$ as in \eqref{eq_sch-op}-\eqref{eq_sch-ic}, for $m=1,\ldots M$. The terms $\| U^{m}(t_m) - W^{m} \|_F^2$ penalize violations of the equality constraint \eqref{eq_con0},
where the coefficient $\mu>0$ is a free parameter that controls the influence of the penalty terms relative to the generalized infidelity. From \eqref{eq_gen-infid}, we find that ${\cal J}_{\cal V} \leq \|U^M\|_F^2/n$. The generalized infidelity and the penalty terms are therefore in approximate balance by choosing 
\begin{align}
    \mu = {\cal O}\left(\frac{1}{n}\right).
\end{align}

We minimize the penalty objective function ${\cal P}$ using a gradient-based quasi-Newton method, where the Hessian is approximated using L-BFGS updates~\cite{nocedal2006numerical}. In order to compute the gradient of ${\cal P}$ with respect to $\alphab$ and $W^m$, we employ an adjoint methodology that is outlined in \S\ref{sec_adjoint-grad}. 
One challenge with a time-parallel implementation of the multiple shooting method is to decide when to terminate the optimization iteration. The conventional (reduced-space) approach is to iterate until the trace infidelity falls below a given threshold. In the context of multiple shooting, this would correspond to evaluating the generalized infidelity \eqref{eq_mod-infid} on the (so-called) roll-out state, which satisfies Schr\"odinger's equation \eqref{eq_schro} under the current control vector. However, evaluating the roll-out state requires time integration over the full time domain $t\in[0,T]$, which necessitates sequential time stepping in each time window. To mitigate this computational bottleneck, we instead use a termination criteria based on an estimate of the generalized infidelity on the roll-out state. This estimate can be evaluated locally within each time window, see \S\ref{sec_opt-stopping} for details.

\subsection{Computing the gradient of the objective function}\label{sec_adjoint-grad}

We start by considering the gradient of the objective function \eqref{eq_penalty-fcn} with respect to the control vector $\alphab$ and consider the intermediate initial conditions to be fixed. 
For each control parameter $\alpha_\ell$, $\ell=1,\dots,d$, the gradient $d {\cal P}/d\alpha_\ell$ depends on $\p U^m(t_m)/\p \alpha_\ell$, due to its implicit dependence on $\alphab$ through the control Hamiltonian in Schr\"odinger's equation. In principle, it can be calculated by first differentiating the state equation \eqref{eq_sch-op} with respect to $\alpha_\ell$, integrating the resulting equation to obtain $\p U^m(t_m)/\p \alpha_\ell$, followed by evaluating $d {\cal P} / d \alpha_\ell$. However, this approach is computationally inefficient when there are more than a few elements in the control vector $\alphab$, as it requires $d$ solutions of the differentiated state equation.
Instead, a more efficient gradient calculation can be derived by seeking stationary points of the associated Lagrangian functional, see e.g.~\cite{nocedal2006numerical}. 
This approach leads to the introduction of the adjoint state variable $\Lambda^m(t)\in\C^{n\times n}$, governed by the adjoint state equation,
\begin{gather}
    \dot{\Lambda}^m(t) + i H(t;\alpha) \Lambda^m(t) = 0,\quad t_{m} > t \geq t_{m-1},\label{eq_lambda-de}\\
    \Lambda^m(t_m) = \begin{cases}
        \frac{\mu}{2}(W^m - U^m),\quad & m=1,\ldots, M-1,\\
        \frac{1}{n^2}\left(\langle {\cal V}, U^M\rangle_F \,{\cal V} - nU^M \right),& m=M,
    \end{cases}\label{eq_terminal}
\end{gather}
where $U^m = U^m(t_m)$.
In each window, this differential equation is solved backwards in time, subject to the terminal conditions \eqref{eq_terminal}. By first solving the state equation \eqref{eq_sch-op}-\eqref{eq_sch-ic}, and then the adjoint state equation \eqref{eq_lambda-de}-\eqref{eq_terminal}, each component of the gradient can be found by evaluating the integral
\begin{align}\label{eq_grad-int}
    \frac{\p {\cal P}}{\p\alpha_\ell} = 
    2\sum_{m=1}^M\mbox{Re}\int_{t_{m-1}}^{t_m} \left\langle \Lambda^m(t), i \frac{\p H(t;\alphab)}{\p \alpha_\ell} U^m(t) \right\rangle_F
    \, dt,\quad\ell = 1,\dots,d,
\end{align}
see~\ref{app_gradient-der} for details.
Computing all components of the gradient thus only requires the solution of two differential equations in each window. 
When using the adjoint formalism, the dominating computational cost of the gradient computation is hence approximately twice the cost of evaluating the objective function itself, but most importantly, it is essentially independent of the dimension of the control vector $d$. 

We proceed by analyzing the gradient with respect to the intermediate initial conditions. Let $S^m = S^m(t_m;\alphab)$, where $\alphab$ is fixed, and recall that $W^0=I$. The terms in the penalty objective function \eqref{eq_penalty-fcn} can then be written as
\begin{align}\label{eq_objective-W}
    P_m = \begin{cases}
        \frac{\mu}{2} \| S^m W^{m-1} - W^m \|_F^2,\quad & m=1,\ldots,M-1,\\
        {\cal J}_{\cal V}(S^M W^{M-1}), & m=M.        
    \end{cases}
\end{align}
Hence, the gradient of the objective function with respect to $W^m$ gets contributions both from ${P}_m$ and ${P}_{m+1}$.
To calculate the gradient of the real-valued objective function $\cal P$ with respect to the complex-valued arguments $W^m$, we first decompose the initial conditions into their real and imaginary parts, $W^m_r = \mbox{Re}(W^m)$, $W^m_i = \mbox{Im}(W^m)$, and separately calculate the corresponding components of the gradient. 
As detailed in~\ref{app_gradient-der}, the gradient of ${\cal P}$ with respect to the initial condition is given by
\begin{align}
    \frac{\p {\cal P}}{\p W^{m}_{x}} &= \mu \left(W^{m}_x - U^{m}_x \right)
    - \Lambda^{m+1}_x(t_m) ,\quad m=1,\ldots,M-1,\label{eq_dP-dW}
\end{align}
for the real ($x=r$) and imaginary ($x=i$) parts of $W^m$.
We emphasize that evaluating this gradient does {\it not} require any additional differential equations to be solved.

\subsection{Termination criteria}\label{sec_opt-stopping} 

{In order to define an appropriate stopping criterion for the optimization iterations of the multiple-shooting quadratic penalty method, we here derive an estimate on the roll-out infidelity for a given control vector.}
Denote the violation of the equality constraint \eqref{eq_con0} in window $m$ by
\begin{align}\label{eq_constraint}
    C^m := S^m W^{m-1} - W^m,\quad \Leftrightarrow\quad W^m = S^m W^{m-1} - C^m,\quad m=1,\ldots,M-1.
\end{align}
Repeated application of \eqref{eq_constraint} gives
\begin{align}
    S^M W^{M-1} = S^M S^{M-1} W^{M-2} - S^M C^{M-1} = \ldots 
    = \left(\Pi_{m=M}^{1} S^m\right) - \sum_{m=1}^{M-1} \left( \Pi_{\ell=M}^{m+1} S^\ell \right) C^m,
\end{align}
or equivalently,
\begin{align}
    U_{ro} &:= \Pi_{m=M}^1 S^m  = U^M + \sum_{m=1}^{M-1} \left( \Pi_{\ell=M}^{m+1} S^\ell \right) C^m.
\end{align}
On the left hand side is the (so called) roll-out state matrix, $U_{ro}$, which follows by solving Schr\"odingers equation in the full time-domain $[0, t_M]$. The first term on the right hand side is the state matrix $U^M = S^M W^{M-1}$, corresponding to only solving Schr\"odingers equation in the final time window, with initial condition $W^{M-1}$. Each term in the sum represents the constraint violation at time $t_m$, forward propagated to time $t_M$.

Let $\vec{U} = \mbox{vec}(U)$ denote the vectorized representation of the matrix $U$. Because the generalized infidelity \eqref{eq_gen-infid} is quadratic in $\vec{U}$ and $Q_v$ is a Hermitian matrix, the roll-out infidelity can be related to the final infidelity and the constraint violations. In \eqref{eq_gen-infid}, we take $\vec{U}_{ro} = \vec{U}^M + \vec{D}$, where $D = \sum_{m=1}^{M-1} \left( \Pi_{\ell=M}^{m+1} S^\ell \right) C^m$, to derive
\begin{align}\label{eq_roll-out-infid-0}
    n J_{\cal V}(\vec{U}_{ro}) = n J_v(\vec{U}^M) + \langle \vec{D} , Q_v\vec{U}^M\rangle + \langle Q_v\vec{U}^M, \vec{D}\rangle + \langle \vec{D}, Q_v \vec{D} \rangle.
\end{align}
The right hand side of this expression can be estimated in terms of the final infidelity and the penalty terms (see~\ref{app_infid-est} for details), resulting in
\begin{align}\label{eq_rollout-infid-est}
    J_{\cal V}(U_{ro}) \leq J_{\cal V}(U^M) + 
    \frac{2}{\sqrt{n}} \sqrt{J_{\cal V}(U^M)} \sum_{m=1}^{M-1} \| {C}^m \|_F 
    + \frac{1}{n}\left( \sum_{m=1}^{M-1} \| C^m \|_F\right)^2.
\end{align}
Here, the norm of $C^m$ can be calculated locally within window $m$, and the term $J_{\cal V}(U^M)$ can be calculated in the final window. The estimate for the roll-out fidelity can thus be evaluated concurrently followed by a reduction operation over scalar quantities.

\section{Parallelization strategy}\label{sec_time-parallel}

\begin{figure}
    \centering
    \includegraphics[width=\textwidth]{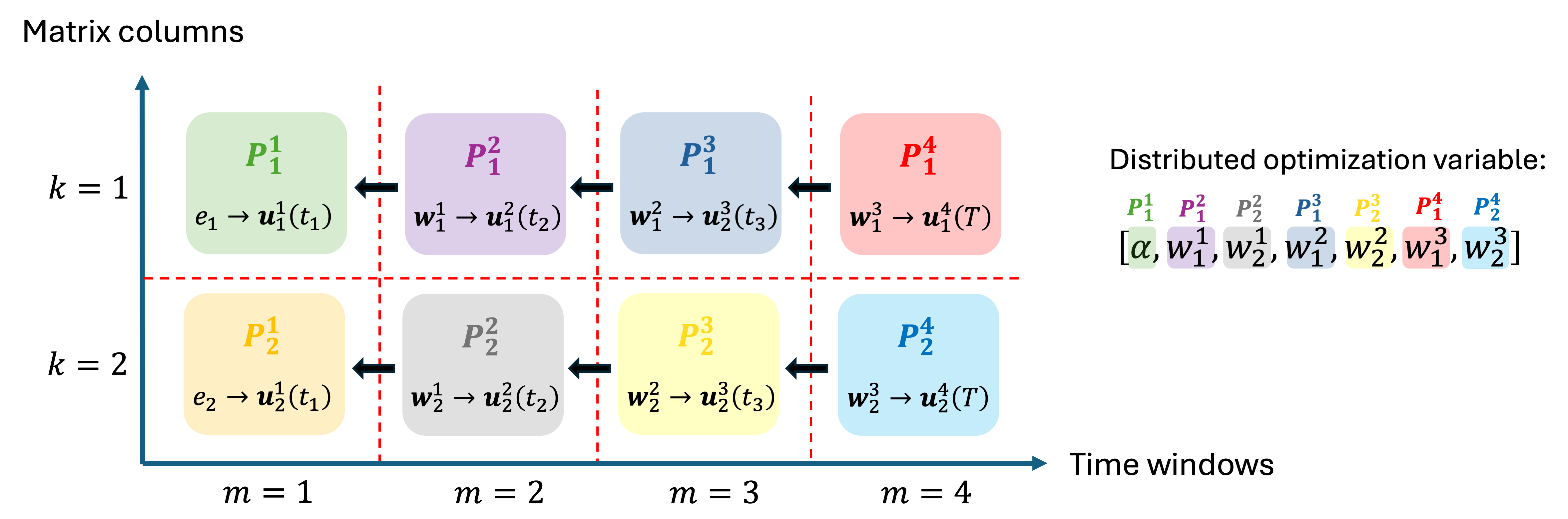}
    \caption{Two-dimensional process grid for distributed initial conditions (columns of the state matrix) and time windows. Each computational process $P_k^m$ solves Schr\"odinger's equation, subject to column $k$ in the initial condition matrix, in time window $m$, by propagating the initial state vector $\wb^{m-1}_{k}$ forward in time to yield $\ub^{m}_{k}(t_{m})=S^m(t_m)\wb_k^{m-1}$. Evaluating the penalty terms $\|\ub^{m}_k(t_m)-\wb_k^{m}\|^2$ requires nearest-neighbor communication along time processes to {obtain} $\wb_k^{m}$ from process $P^{m+1}_k$.} 
    \label{fig:2Dprocessorgrid}
\end{figure}

The reformulation of the optimal quantum control problem into a time-windowed multiple-shooting problem enables us to parallelize the evaluation of the objective function (and its gradient) along the time domain. To this end, we apply a distributed-memory parallelization strategy using the message-passing interface (MPI)~\cite{mpibook}, where each process is assigned to one computational processor (core) in the hardware. Solving Schr\"odinger's equation in each time window can be performed concurrently using $M$ computational processes along the time domain, each propagating the state matrix $U^m(t)$ from the intermediate initial state $W^{m-1}$ at time $t_{m-1}$ to yield $U^{m}(t_m)$. In addition, the matrix-valued Schr\"odinger equation naturally parallelizes over the columns of each initial condition matrix $W^{m} =: [\wb^{m}_1,\dots \wb^{m}_{n}]$, such  that each column can be evolved independently, yielding an orthogonal opportunity for computational concurrency along the state columns. Together, we hence define a 2-dimensional grid of computational processes, parallelizing over the $M$ time windows as well as the $n$ columns of the initial states, $P^{m=1,\dots M}\times P_{k=1,\dots, n}$. Thus, process $P^m_k$ computes the $k$-th column of $U^m$ by solving Schr\"odinger's equation in window $m$, subject to the initial state $\wb^{m-1}_{k}$. Figure \ref{fig:2Dprocessorgrid} exemplifies a fully distributed two-dimensional process grid for $M=4$ time windows and $n=2$ state columns, as well as the corresponding distributed storage of the optimization variables. Note that the initial condition in the first window is fixed at $W^{0}=I$ and hence is not part of the distributed optimization variables. Instead, we assign the first process $P_1^1$ to hold the control vector $\alphab$. Because the B-spline basis functions have compact support in time, we note that only a subset of the control vector is needed to solve Schr\"odinger's equation in each time window. Hence, in principle, $\alphab$ could also be distributed and stored locally alongside $\wb^m_k$.

In the distributed memory model used by MPI, each core only has access to a subset of all variables, as exemplified in Figure \ref{fig:2Dprocessorgrid}; access to other variables is provided by explicitly calling message passing functions.
In order to evaluate the penalty objective function \eqref{eq_penalty-fcn}, the following communication scheme is applied. Since the control vector $\alphab$ is needed in all time windows, but stored locally on the first process, we first broadcast $\alphab$ from $P_1^1$ to all other processes. To evaluate the local contribution to the penalty term, $\|\ub_k^{m}(t_{m}) - \wb_{k}^{m}\|^2$, for each column $k$ and window $m$, only nearest-neighbor communication along the time axis is needed. Efficient communication is achieved through non-blocking MPI functionality in the following way: Each processor $P_k^m$ first initiates a non-blocking \texttt{MPI\_Isend} of its column $\wb^{m-1}_k$ to its left neighbor in time, $P_{k}^{m-1}$. Then, $P_k^m$ computes $\ub_k^{m}(t_m)$ by solving Schr\"odinger's equation in its time window, before calling a non-blocking \texttt{MPI\_IReceive} from its right neighbor, $P^{m+1}_k$, to receive $\wb_{m}^k$ such that $\|\ub^{m}_k(t_m)-\wb^{m}_k\|^2$ can be evaluated on this process. The last time-processes, $P^M_k$, evaluate their contribution to the generalized infidelity at time $t_M$. The penalty objective function is then assembled by summing up the contributions from each time- and column-process through an \texttt{MPI\_Allreduce}, gathering the objective function value and distributing it to all processes.

The gradient computation follows a similar communication pattern. Each process $P^m_k$ receives $\wb^{m-1}_k$ from its left neighbor in the time axis, and solves the adjoint state equations backwards in its time window, starting from the terminal condition, $\lambdab^m_k(t_{m})$, that is stored on this process. The backwards time stepping is performed concurrently in each time window and for each column in $\Lambda^m$, yielding the initial adjoint states $\lambdab^m_k(t_{m-1})$ to assemble the locally stored gradients $\partial {\cal P} / \partial {\wb^{m-1}_k}$ as in \eqref{eq_dP-dW}, as well as its contribution to the gradient with respect to $\alphab$ as in $\eqref{eq_grad-int}$. Assembling $\partial {\cal P} / \partial {\alphab}$  then requires an additional communication step that sums the contributions from each time window and column onto $P_1^1$. 

\section{Numerical results}\label{sec_num-exp}

To demonstrate the performance and parallel scalability of the time-parallel multiple-shooting quantum control problem, we here consider the $n$-dimensional Quantum Fourier Transformation (QFT) represented by the unitary target matrix $V_{QFT_n} \in \mathbb{C}^{n\times n}$. Using zero-based indexing, the elements of the matrix are given by:
\begin{align}
    \{ V_{QFT_n} \}_{jk} = \frac{1}{\sqrt{n}} \,\kappa^{jk},\quad 0 \leq j,k \leq n-1,\quad \kappa = e^{i 2\pi/n}.
\end{align}
We consider $n\in\{4,8,16\}$ corresponding to $q=2,3$ and $4$ qubits, respectively, which are coupled in a linear chain with system parameters given in Table \ref{tab:systemparams}, based on the Hamiltonian model in \eqref{eq_sys-ham}-\eqref{eq_ctrl-ham}, with rotation frequency $\omega^{rot} = 1/q \sum_{j=1}^q \omega_j$.

\begin{table*}
 \centering
 \begin{tabular}{@ { } l | l l l @ { }}
      Testcase   &   $\omega_j/2\pi$ [GHz] & $J_{j, j+1}/2\pi$ [MHz] & $\Omega_j/2\pi$ [MHz] \\
   \hline
      QFT-4      & 1: $5.18$ & $1\leftrightarrow 2$: 5.0 &$[-30.41, 30.41]$ \\
      (2 qubits) & 2: $5.12$ &     &$[-30.41,30.41]$ \\
    \hline  
      QFT-8      & 1: $5.18$ & $1\leftrightarrow 2$: 5.0& $ [0, -60.4]$ \\
      (3 qubits) & 2: $5.12$ & $2\leftrightarrow 3$: 5.0& $ [60.4,0, -60.4]$ \\
                 & 3: $5.06$ &    & $ [60.4,0]$ \\
    \hline 
      QFT-16     & 1: $5.18$ & $1\leftrightarrow 2$: 5.0 & $[-30.0, -90.41]$ \\
      (4 qubits) & 2: $5.12$ & $2\leftrightarrow 3$: 5.0 & $[30.0, -30.0, -90.41]$ \\
                 & 3: $5.06$ & $3\leftrightarrow 4$: 5.0 & $[90.41, 30.0, -30.0]$ \\
                 & 4: $5.0$  &                           & $[90.41, -30.0]$ \\
    \hline
\end{tabular}
  \caption{Transition frequencies, dipole-dipole coupling strengths and carrier wave frequencies in the Hamiltonian model \eqref{eq_sys-ham}-\eqref{eq_ctrl-ham} for $q=2,3,4$ qubits that are coupled in a linear chain.}\label{tab:systemparams}
\end{table*}

We implement the multiple-shooting formulation in the Quandary software \cite{quandaryGithub}, which provides a structure preserving and time-reversible integration scheme to solve Schr\"odinger's equation in any given time domain using the implicit midpoint rule, as well as backwards time-stepping based on the discrete adjoint approach to calculate gradients, hence employing the "discretize-then-optimize" paradigm that allows the gradient to be calculated exactly in the time-discretized setting, compare \cite{gunther2021quandary} for details. The adjoint states are utilized to evaluate the gradient of the multiple-shooting objective function with respect to the control parameters $\alphab$ and the initial condition states $W^m$ in a real-valued formalism.
We parameterize the control pulses $d_j(t, \alphab)$ for each qubit $j$ using a B-spline knot-spacing of $\Delta_\tau = 3$ ns, resulting in $d_1 = \lceil T/\Delta_{\tau}\rceil +2$ B-spline coefficients per carrier wave frequency and hence total of $d= 2 d_1 \sum_j N_f(j)$ real-valued control parameters $\alphab\in \R^d$, where the total number of carrier frequencies, $N_{f}$, depends on the number of qubits and their connectivity, as shown in Table \ref{tab:systemparams}. The multiple-shooting optimal control formalism further considers the real and imaginary parts of the intermediate states $W^m, m=1,\dots M$ as additional optimization variables, resulting in an optimization vector of size $d + 2n^2(M-1)$. 

\begin{figure}
    \centering
    \includegraphics[width=0.49\textwidth]{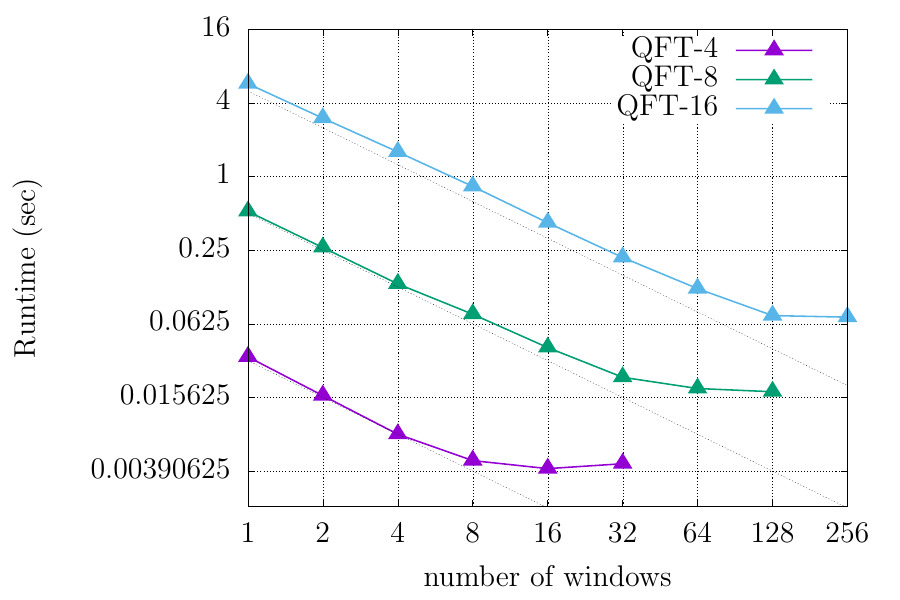}
    \includegraphics[width=0.49\textwidth]{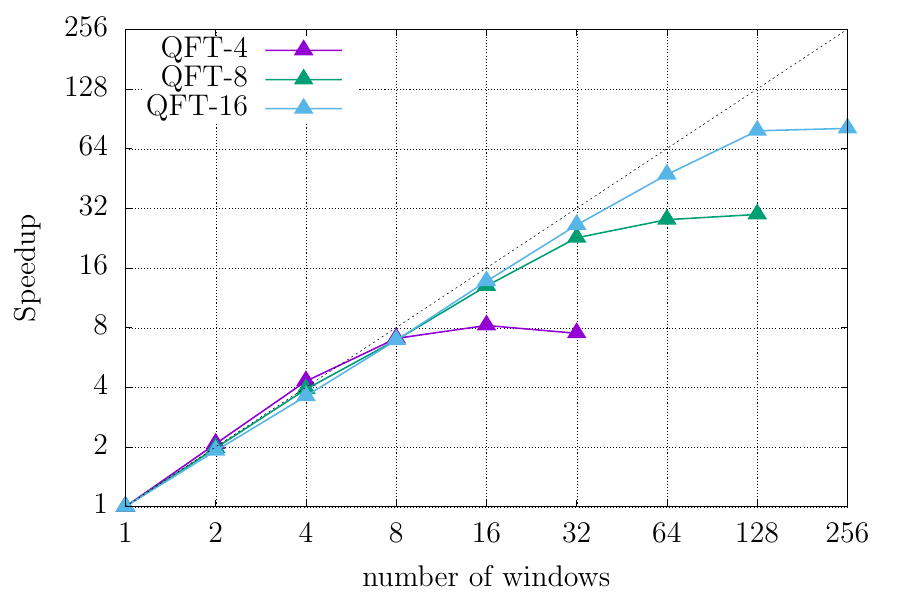}\\
    \caption{Parallel scaling study for one gradient evaluation: Runtime (left) and parallel speedup (right) of the multiple-shooting gradient evaluation over single-shooting reference for the QFT gate on 2, 3 and 4 qubits ($n=4,8,16$).}
    \label{fig:QFT_grad_scaling}
\end{figure}

In order to investigate the parallel scalability of the multiple-shooting algorithm, we first measure run times for one single gradient evaluation of the penalty objective function, and compare it to the run time for the standard reduced-space gradient with respect to $\alphab$ only, i.e., the single-window case with $M=1$. 
We distribute the optimization variables onto the 2-dimensional process grid as described in Section \ref{sec_time-parallel} and concurrently execute the computation over the $n$ columns of the state matrix and the $M$ time-windows, such that the total number of computational processes used for each data point in the following plots equals $nM$. 
Figure \ref{fig:QFT_grad_scaling} shows a parallel scaling study performed on LLNL's High-Performance Cluster "Dane"\footnote{LLNL Livermore Computing "Dane" platform: Intel Sapphire Rapids 2.0GHz, 112 cores per node, 256 RAM per node, Cornelis Networks Interconnect}, measuring the computational runtime for one gradient evaluation over increasing numbers of time windows, for the QFT gate on 2, 3 and 4 qubits ($n=4,8,16$).
As expected, we observe excellent parallel scalability with respect to both the time-window distribution as well as the columns in the initial condition matrices. A maximum speedup in runtime for one gradient evaluation of about $80\times$ is observed for the 4-qubit test case when using 256 time-windows and 16 column processes, see Table~\ref{tab:testcases}. Note that the total number of time-steps $N_T$ for each test case was chosen such that the absolute error in the infidelity due to time-stepping is less than $10^{-6}$. 

\begin{table}
  \center
  \begin{tabular}{@ { } llllr @ { }}
      \toprule
        Test case   &$\#$qubits&  Gate duration & Number of time-steps & Max. Speedup (Gradient) \\
	 \midrule
        QFT-4 & 2 & $190$ns & $N_T= 2,252$     & 8$\times$ on 16 time-processes\\
        QFT-8 & 3 & $500$ns & $N_T= 19,806$    & 31$\times$ on 128 time-processes\\
        QFT-16 & 4 & $900$ns & $N_T= 106,072$   & 80$\times$ on 256 time-processes\\
     \bottomrule
  \end{tabular}
  \caption{Gate durations of the QFT-4,-8, and -16 testcase, the number of total time steps, as well as the maximum measured runtime speedup for one gradient evaluation in the multiple-shooting formulation compared to the single-window reference case.}
  \label{tab:testcases}
\end{table}

Next, we investigate the parallel scalability of the multiple-shooting optimization algorithm when increasing the numbers of time-windows and corresponding computational processes. As a reference point, we solve the reduced-space optimization problem (one time window), which performs standard L-BFGS iterations on the control parameters $\alphab$, as implemented in the Quandary software, based upon the PETSc/TAO library~\cite{petsc-web-page}. 
Throughout the optimization, we impose box-constraints on the control parameters such that the resulting optimized pulse amplitudes are below $25$MHz.
To regularize the optimization problem, we add a Tikhonov regularization term to the objective function with coefficient $\gamma_{tik} = 10^{-3}/d$, as well as a penalty integral term to minimize the control pulse energy with coefficient $\gamma_{energy} = 10^{-3}$, see the Quandary user's guide~\cite{quandaryGithub} for details. 
The initial guess for each element in the control vector $\alphab$ is generated using a random variable, uniformly distributed in the range $[-10, 10]$ MHz. 

For the time-parallel multiple-shooting optimization strategy, we fix the quadratic penalty coefficient in \eqref{eq_penalty-fcn} at $\mu = 2/n$. To initialize the intermediate state optimization variables $W^m$, we roll out the state matrices using the initial control vector $\alphab$, by solving Schr\"odingers equation \eqref{eq_schro} and recording the state at the time-window boundaries. Hence, all equality constraints are initially satisfied, but the initial infidelity is generally far from zero. 
The termination criteria for the multiple-shooting optimization iteration is chosen such that the estimated roll-out infidelity based on \eqref{eq_rollout-infid-est} is less than $10^{-3}$. 
In order to balance the contributions from the control vector and the intermediate state matrices in the gradient, it is suggested in \cite{JanMey-24} to scale the intermediate state matrices within the optimization, $\widetilde{W}^m = \sigma W^m$, for a $\sigma>0$. 
Following this idea, we here tune the scaling factor by scanning over a range $\sigma \in [0.05:0.15]$ and pick the scale factor that yields the smallest number of optimization iterations. We note that further research is needed to determine the optimal choice of this scale factor.

\begin{figure}
    \centering
    \includegraphics[width=.65\textwidth]{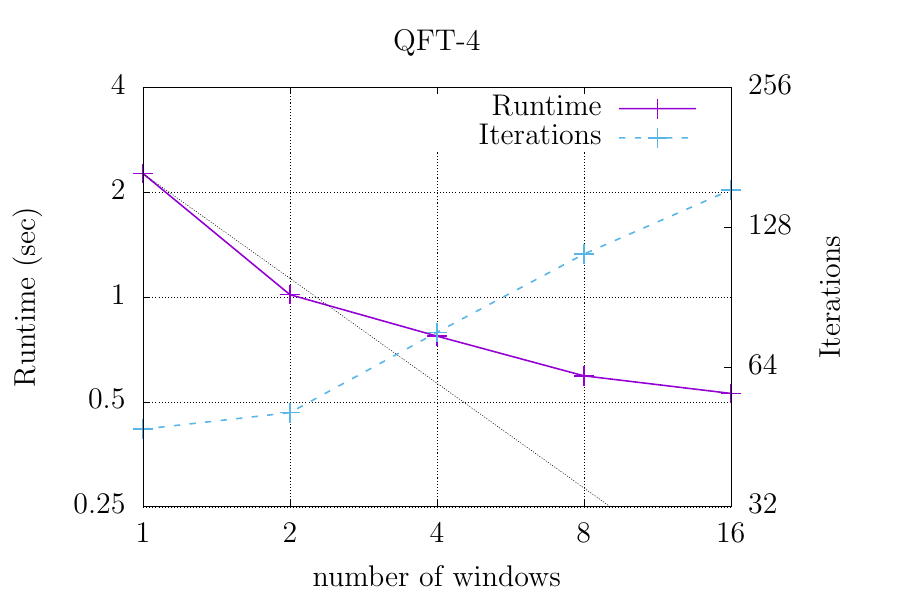}\\
    \includegraphics[width=.65\textwidth]{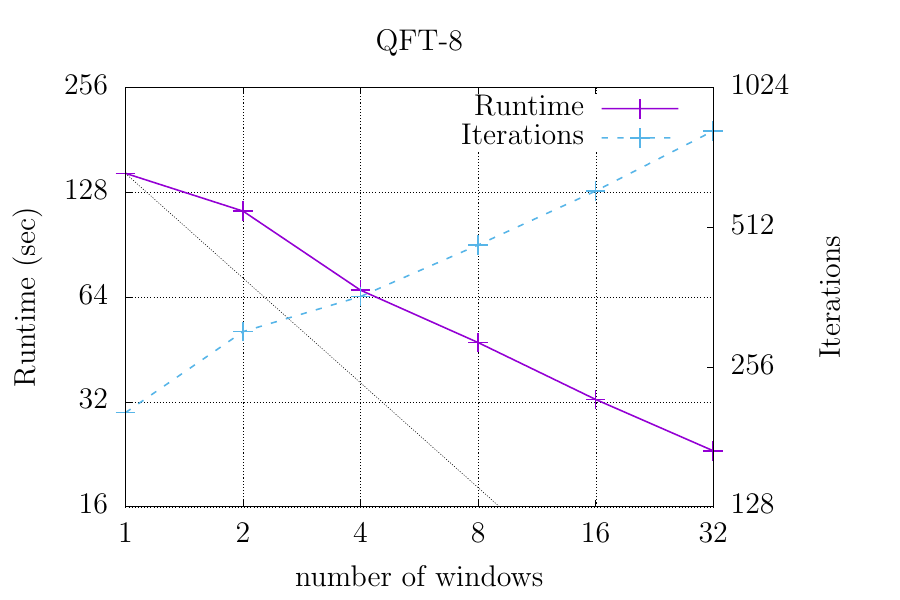}\\
    \includegraphics[width=.65\textwidth]{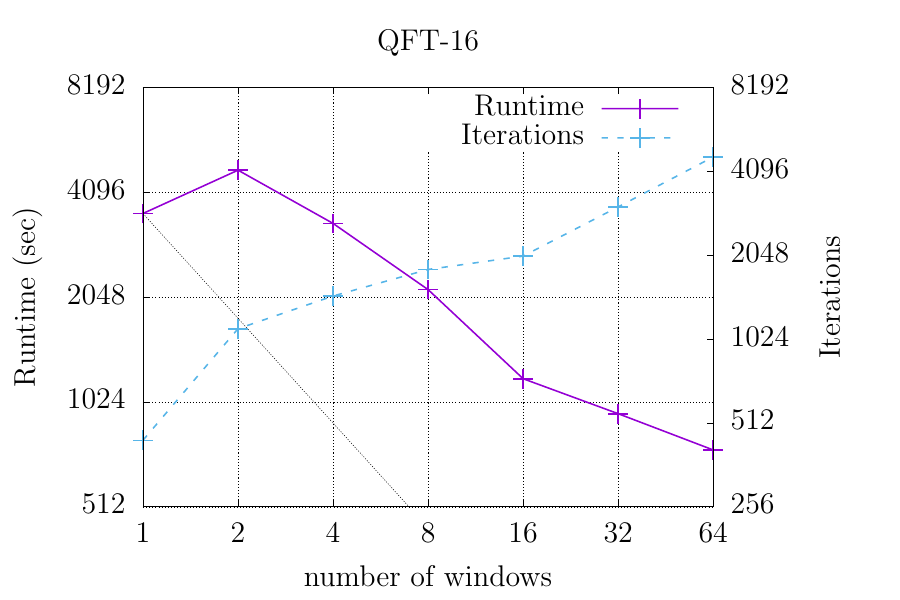}\\
    \caption{Parallel scaling study of the multiple-shooting optimization for the QFT gate on $q=2,3$, and $4$ qubits ($n=4,8,16$): Runtime (solid lines) and number of optimization iterations (dashed lines) for increasing numbers of time windows $M$. For each case, the total number of distributed computing tasks is $nM$.}
    \label{fig:optimresults}
\end{figure}

\begin{figure}
    \centering
    \includegraphics[width=0.7\textwidth]{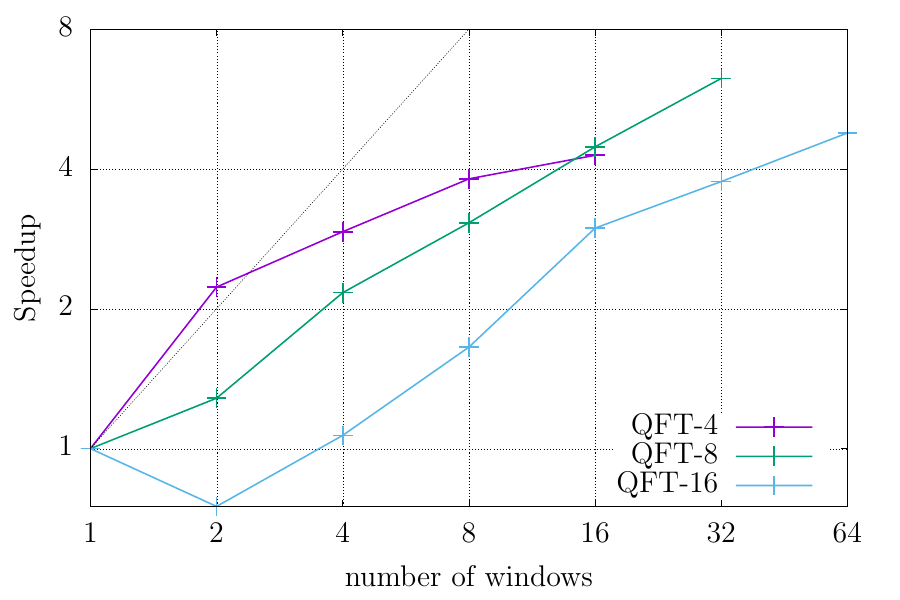}
    \caption{Runtime speedups of the multiple-shooting optimization over the reference single-window optimization for the QFT gate on $q=2,3,4$ qubits ($n=4,8,16$).}
    \label{fig:optimspeedup}
\end{figure}

Parallel scaling results for the multiple-shooting optimization as well as the single-window reference optimization are presented in Figure \ref{fig:optimresults}. Here, we plot both the resulting overall runtime of the optimization (left axis, solid lines) performed on $nM$ computational processes, as well as the number of optimization iterations needed for convergence (right axis, dashed lines), for the QFT$_n$ test cases on 2,3  and 4 qubits ($n=4,8,16$). First, we note that the number of iterations needed increases monotonically with the number of time windows $M$, while the jump from the single-window reference optimization to multiple-window optimization can be significant for the larger test cases (e.g., it doubles for the 4-qubit case from $M=1$ to $M=2$). We attribute this increase to poor performance of the quadratic penalty method for solving the constrained multiple-shooting optimization problem. We note that a more elaborate optimization strategy, such as, for example, the primal-dual interior point method in the Ipopt \cite{Wachter2006} solver could potentially improve the convergence. Nevertheless, despite this increase in optimization iterations, Figure \ref{fig:optimresults} shows that significant parallel speedup can be achieved through the time-parallel multiple-shooting formulation due to significant speedup of each gradient evaluation, which is parallelized over $M$ time-windows, as demonstrated in Figure \ref{fig:QFT_grad_scaling}. We observe that the parallel scaling of the multiple-shooting optimization generally improves for larger test cases, showing nearly perfect scaling on the QFT-16 case between $M=2$ to $M=16$. The run time speedups of the multiple-shooting optimization over the single-shooting formulation are plotted in Figure \ref{fig:optimspeedup}, where the best speedups are summarized in Table \ref{tab:optimresults}, alongside the corresponding roll-out gate infidelities. 
While the parallel efficiencies\footnote{Parallel efficiency is defined as the ratio of the runtime for single-shooting optimization divided by the product of the runtime for multiple-shooting optimization and the number of computational processes.} of those results are less than perfect (for example, the parallel efficiency for the 6.3x speedup for the QFT-8 optimization on 32 time-processes is~$~20\%$) we note that the resulting speedup in runtime is significant, reducing the runtime for the 3-qubit gate optimization from 145 seconds to 23 seconds, and for the 4-qubit gate from 59 minutes to 12 minutes. Of significant practical importance is also that the 2-qubit optimal control problem can be solved in only 0.5 seconds, when running on a total of 64 processes (16 time-processes and 4 column-processes). Hence, utilizing available HPC resources at this scale can significantly advance quantum control techniques and has the potential to yield higher-fidelity quantum controls in real-time that are ready to be integrated in modern quantum compilers.

\begin{table}
  \center
  \begin{tabular}{@ { } l | lllr @ { }}
      Test case & Windows $(M)$  & Gate infidelity & Runtime & Speedup  \\ 
	 \hline
        QFT-4   & 1  &     2.37e-04  & {2.3 sec} & --  \\
                & 16  &    1.49e-04  & {0.5 sec} & 4.3x \\
        \hline
        QFT-8  & 1  & 2.44e-04 & {145 sec} & -- \\
               & 32 & 8.86e-05 & {23 sec} & 6.3x \\
        \hline
        QFT-16 & 1  & 1.59e-04 & {59 min} & --\\
               & 64 & 6.33e-05 & {12 min} & 4.8x \\
    \hline
  \end{tabular}
  \caption{Optimized gate infidelity and maximum observed speedup in runtime for the multiple-shooting optimization compared to the single-window reference.}
  \label{tab:optimresults}
\end{table}

\begin{figure}
    \centering
    \includegraphics[width=0.7\textwidth]{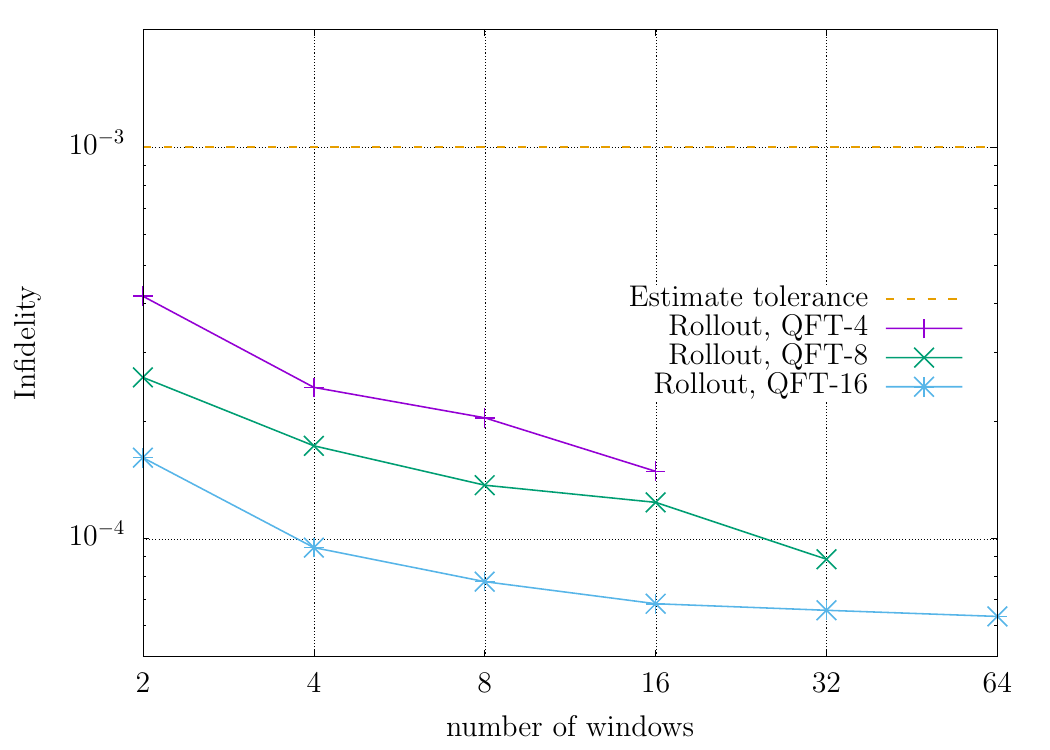}
    \caption{Optimized rollout infidelities at the last multiple-shooting optimization iteration.}
    \label{fig:rolloutestimate}
\end{figure}

We note that reported speedups depend on the choice of the stopping criterion of the optimization iterations. In particular, the multiple-shooting optimization iterations are stopped when the estimate on the fidelity drops below $10^{-3}$, where Figure \ref{fig:rolloutestimate} plots the resulting rollout gate infidelities at the last iteration of the multiple-shooting optimization. We observe that the estimate gets less and less sharp with increasing numbers of qubits as well with the number of time-windows used for each test case, such that the achieved rollout infidelities at the optimal point can be an order of magnitude smaller than the estimate (the stopping criterion). To ensure a reasonably fair comparisons of the multiple-shooting run-times with the reduced-space single-shooting reference optimization, we stop the single-shooting optimization iterations when its infidelity falls below the roll-out infidelity achieved for the $M=2$ window test cases (the left-most data points in Figure \ref{fig:rolloutestimate}). Note that this choice favours the single-shooting reference optimization, since the roll-out fidelity's for $M>2$ will be even smaller that those achieved in the single-shooting case, and hence requires more iterations to achieve the same level of infidelity.

\section{Conclusions}\label{sec_conclusions}

In this paper we derived a multiple-shooting formalism for solving the quantum optimal control problem, focusing on inducing unitary gate transformations in closed quantum systems. 
The multiple-shooting algorithm is based on splitting the time-domain into segments (windows) and introducing the state matrices at the beginning of each window as additional optimization variables. As a result, time-stepping in each window can be performed concurrently, allowing significant acceleration through parallel processing along the time domain on an HPC system. During the optimization iteration, the state matrices representing the initial conditions in each window may become non-unitary. For this reason, we have broadened the definition of the conventional gate trace infidelity, resulting in a generalized infidelity that is a non-negative and convex function of a general complex state matrix. 
Continuity across time window boundaries is enforced through equality constraints. Here we solve the constrained optimization problem using the quadratic penalty method.
To calculate the gradient of the objective function with respect to the controls and the intermediate initial conditions efficiently, we have shown that the adjoint state, which is used to calculate the gradient with respect to the control vector, can also be used to calculate the gradient with respect to the state at the beginning of each time window. As a result, all components of the gradient can be evaluated by solving one state equation and one adjoint state equation in each window, at a cost that essentially is independent of the number of optimization variables. 
We derived an estimate for the roll-out infidelity based on the violations of the equality constraints across window boundaries that serves as a stopping criterion for the multiple-shooting algorithm, guaranteeing that the roll-out infidelity meets a prescribed tolerance without explicit evaluation.

An efficient time-parallelization strategy for concurrent evaluation of the objective and its gradient on multiple distributed-memory compute processes was also presented. This strategy has been implemented in the Quandary software~\cite{quandaryGithub}, which was used to evaluate the parallel scaling performance in quantum systems with 2, 3, and 4 coupled qubits, subject to Quantum Fourier Transform (QFT) gates. 
We observed significant acceleration for the time-parallel evaluation of the objective and its gradient, with nearly perfect scaling with respect to increasing numbers of time windows and corresponding computational processes. However, the number of optimization iterations in the quadratic penalty method was found to grow with the number of time windows, hampering the overall speedup in runtime for the multiple-shooting optimization.
Nevertheless, using the proposed method, the 2-qubit optimal control problem can be solved in only 0.5 seconds when running on a total of 64 compute cores. 
Hence, utilizing available classical compute resources can significantly advance quantum control techniques and has the potential to yield higher-fidelity quantum controls in real-time that are ready to be integrated in modern quantum compilers.
For larger gates, we observed a maximum runtime speedup of 6.3x and 4.8x for the 3- and 4-qubit QFT gates, when compared to standard reduced-space optimization.
To fully reap the benefits of the time-parallel multiple shooting method, more research is needed to identify more effective algorithms for solving the constrained multiple-shooting optimization problem, such as, for example, augmented Lagrangian methods, sequential quadratic programming (SQP) techniques, or primal-dual interior point approaches.

\section*{Acknowledgements}
The authors gratefully acknowledge partial support from the Department of Energy, Office of Advanced Scientific Computing Research, under project TEAM (Tough Errors Are no Match), grant number 2020-LLNL-SCW1683.1. The authors are also grateful for partial support under the Laboratory Directed Research and Development program at LLNL (grant number 23-ERD-038). 

\appendix
\section{Estimating the roll-out infidelity}\label{app_infid-est}

Starting from \eqref{eq_roll-out-infid-0}, we proceed by estimating the terms in the right hand side. First, consider
\begin{align}
    Q_v \vec{U}^M = \left( I - \frac{1}{n} \vec{\cal V} \vec{\cal V}^\dagger \right) \vec{U}^M = \vec{U}^M - \frac{\gamma_M}{n} \vec{\cal V},\quad 
    \gamma_M = \langle \vec{\cal V}, \vec{U}^M \rangle = \langle {\cal V}, U^M \rangle_F.
\end{align}
Therefore,
\begin{align}\label{eq_a2}
    \langle \vec{D}, Q_v\vec{U}^M \rangle = \sum_{m=1}^{M-1} \langle \widetilde{C}^m, U^M - \frac{\gamma_M}{n} {\cal V} \rangle_F.
\end{align}
where $\widetilde{C}^m = \left( \Pi_{\ell=M}^{m+1} S^\ell \right) C^m$ is the forward propagated constraint violation of window $m$.
Because the solution operators are unitary, $\| \widetilde{C}^m \|_F = \| C^m \|_F$, and Cauchy-Schwartz inequality gives the estimate
\begin{align}\label{eq_a3}
    \left| \langle \widetilde{C}^m, U^M - \frac{\gamma_M}{n} {\cal V} \rangle_F \right| \leq 
    \| {C}^m \|_F \left\| U^M - \frac{\gamma_M}{n} {\cal V} \right\|_F.
\end{align}
The factor $\| U^M - \frac{\gamma_M}{n} V\|_F$ can be analyzed further. We have
\begin{multline}\label{eq_a4}
    \left\| U^M - \frac{\gamma_M}{n}{\cal V} \right\|_F^2 = \left\langle  U^M - \frac{\gamma_M}{n}{\cal V},  U^M - \frac{\gamma_M}{n}{\cal V} \right\rangle_F \\
    = \langle U^M, U^M \rangle_F - \frac{\gamma_M^*}{n} \langle {\cal V}, U^M\rangle_F - \frac{\gamma_M}{n} \langle U^M, {\cal V}\rangle_F +  \frac{|\gamma_M|^2}{n^2} \langle {\cal V}, {\cal V} \rangle_F \\
    = \langle U^M, U^M \rangle_F - \frac{2}{n}\langle U^M, {\cal V}\rangle_F \langle {\cal V}, U^M\rangle_F + \frac{1}{n} \langle U^M, {\cal V}\rangle_F \langle {\cal V}, U^M\rangle_F \\
    = n J_{\cal V}(U^M),
\end{multline}
because ${\cal V}$ is unitary. By combining \eqref{eq_a2}-\eqref{eq_a4},
\begin{align}\label{eq_term1}
    \left| \langle \vec{D} , Q_v\vec{U}^M\rangle + \langle Q_v\vec{U}^M, \vec{D} \rangle \right| 
    \leq 2 \sqrt{n J_{\cal V}(U^M)} \sum_{m=1}^{M-1} \| {C}^m \|_F.
\end{align}

To estimate the last term on the right hand side of \eqref{eq_roll-out-infid-0}, we note that
\begin{align}
    Q_v \vec{D} = \left( I - \frac{1}{n} \vec{\cal V} \vec{\cal V}^\dagger \right) \vec{D} = \vec{D} - \frac{\delta_M}{n} \vec{\cal V},\quad \delta_M = \langle \vec{\cal V}, \vec{D} \rangle = \langle {\cal V}, D \rangle_F,
\end{align}
and
\begin{align}\label{eq_term2}
    \langle \vec{D}, Q_v \vec{D}\rangle = \langle \vec{D}, \vec{D} \rangle - \frac{|\delta_M|^2}{n}
    \leq \| D \|_F^2 
    = 
    \left\| \sum_{m=1}^{M-1} \widetilde{C}^m \right\|_F^2 
    \leq
    \left( \sum_{m=1}^{M-1} \| C^m \|_F\right)^2,
\end{align}
because the Frobenius matrix norm is sub-additive (satisfies the triangle inequality). Applying the estimates \eqref{eq_term1} and \eqref{eq_term2} to \eqref{eq_roll-out-infid-0} enables the roll-out infidelity to be estimated in terms of the final infidelity and the constraint violations, as stated in \eqref{eq_rollout-infid-est}.

\section{Derivation of the adjoint gradient}\label{app_gradient-der}

\paragraph{Gradient with respect to $\alphab$}
The penalty objective function \eqref{eq_penalty-fcn} can be written as 
\begin{align}
    {\cal P}(\alphab,W^1,\ldots,W^M) &=  \sum_{m=1}^{M} P_m,
\end{align}
where
\begin{align}\label{eq_penalty-win-j}
    P_m = \begin{cases}
        \frac{\mu}{2}\| U^m - W^m \|^2_F,\quad & m=1,\ldots,M-1,\\
        \frac{1}{n^2}\left( n\| U^M\|_F^2 - \langle U^M, {\cal V}\rangle_F \langle {\cal V}, U^M\rangle_F \right) ,& m = M,
    \end{cases}
\end{align}
with $U^m := U^m(t_m,\alphab)$. The chain rule of differentiation gives
\begin{align}
    \frac{\p P_m}{\p \alpha_\ell} = 
    \begin{cases}
        \frac{\mu}{2}\left(
        \left\langle U^m - W^m, \frac{\p U^m}{\p \alpha_\ell}\right\rangle_F + \left\langle \frac{\p U^m}{\p \alpha_\ell}, U^m - W^m\right\rangle_F \right),\quad & m\in[1,M-1],\\ 
        \frac{1}{n^2}\left(\left\langle n U^M - \gamma_M {\cal V}, \frac{\p U^M}{\p \alpha_\ell} \right\rangle_F +  
        \left\langle \frac{\p U^M}{\p \alpha_\ell}, n U^M - \gamma_M{\cal V}\right\rangle_F \right),\quad & m=M,
    \end{cases}
\end{align}
for $\ell=1,\dots,d$, where $\gamma_M = \langle{\cal V}, U^M\rangle_F$.
By differentiating \eqref{eq_sch-op}-\eqref{eq_sch-ic} with respect to $\alpha_\ell$, we find that $V^m_\ell(t) := \p U^m/\p \alpha_\ell(t)$ satisfies the differential equation
\begin{align}
    \dot V^m_\ell(t) + iH(t; \alphab) V^m_\ell(t) &= F^m_\ell(t), \quad t\in(t_{m-1},t_m], \label{eq_u-alpha-de}\\
    V^m_\ell(t_{m-1}) &= 0,
\end{align}
where the forcing function satisfies
\begin{align}\label{eq_schrod-forcing}
    F^m_\ell(t) = -i\frac{\p H(t;\alphab)}{\p \alpha_\ell}U^m(t).
\end{align}
However, calculating $d {\cal P} / d \alpha_\ell$ by first solving for $V^m_\ell(t_m)$, for each $\ell=1,\ldots, d$, is computationally inefficient when $d\gg 1$. Instead we introduce the adjoint state variable $\Lambda^m(t)\in\C^{n\times n}$, which is chosen to satisfy the differential equation
\begin{align}
    \dot{\Lambda}^m(t) + i H(t;\alpha) \Lambda^m(t) &= 0,\quad t_{m} > t \geq t_{m-1},\label{eq_lambda-de-a}\\
    \Lambda^m(t_m) &= \Xi_m,
\end{align}
for $m=1,\ldots,M$. This is the adjoint state equation, which is solved backwards in time subject to a terminal condition $\Xi_m$. To derive a formula for $\p {\cal P}/\p\alpha_\ell$, we consider time window $m$ and study the expression
\begin{align}
    A_m := \int_{t_{m-1}}^{t_m} \left\langle \dot \Lambda^m + iH\Lambda^m, V^m_\ell\right\rangle_F + \left\langle V^m_\ell, \dot \Lambda^m + iH\Lambda^m\right\rangle_F\, dt.
\end{align}
After integration by parts and noting that $\dot V^m_\ell + iH V^m_\ell = F^m_\ell$, and that $A_m=0$ because $\Lambda^m(t)$ satisfies \eqref{eq_lambda-de-a},
\begin{align}\label{eq_magical-rel}
    0 = \left\langle \Lambda^m(t_m) , V^m_\ell(t_m) \right\rangle_F + \left\langle V^m_\ell(t_m), \Lambda^m(t_m) \right\rangle_F -
    \int_{t_{m-1}}^{t_m} \left\langle \Lambda^m, F^m_\ell \right\rangle_F + \left\langle F^m_\ell, \Lambda^m \right\rangle_F.
\end{align}
To evaluate the boundary terms in the above expression, we choose the terminal condition for $\Lambda^m$ to be
\begin{align}\label{eq_terminal-app}
    \Xi_m =\begin{cases}
        \frac{\mu}{2}(W^m - U^m(t_m)),\quad & m=1,\ldots, M-1,\\
        \frac{1}{n^2}\left(\gamma_M{\cal V} - nU^M(t_M) \right),& m=M.
    \end{cases}
\end{align}
By substituting $\Lambda^m(t_m) = \Xi_m$ into \eqref{eq_magical-rel}, and substituting $F^m_\ell$ from \eqref{eq_schrod-forcing}, we arrive at \eqref{eq_grad-int}.

\paragraph{Gradient with respect to initial conditions}
To derive the formulae for the gradient with respect to the intermediate initial conditions, we start from $P_m$ as defined in \eqref{eq_objective-W} and use the notation from \S~\ref{sec_adjoint-grad}. The gradient  of $P_m$ with respect to $W^m$ satisfies
\begin{align}
    \frac{\p P_m}{\p W^m_{x,pq}} = \mu\,\left\langle -\eb_p \eb_q^\dagger, U^m_x - W^m_x\right\rangle_F = \mu \left\{ W^m_{x} - U^m_x \right\}_{pq},\quad m=1,\ldots,M-1,
\end{align}
for the real ($x=r$) and imaginary parts ($x=i$), where $U^m = S^m W^{m-1}$ and $W_{pq}$ denotes the element in row $p$ and column $q$ of the matrix $W$. Furthermore, $\eb_j$ denotes the $j$-th canonical unit vector.
By differentiating \eqref{eq_objective-W} with respect to the real and imaginary parts of $W^{m-1}$, 
\begin{align}\label{eq_dT-dW-j-1}
    \frac{\p P_m}{\p W^{m-1}_{x,pq}} = \mu \left\langle \eb_p\eb_q^\dagger, W^{m-1}_x - \widetilde{U}^{m-1}_{x}\right\rangle_F
    %
    %
    = \mu\, \left\{ W^{m-1}_x - \widetilde{U}^{m-1}_{x} \right\}_{pq},\quad m=2,\ldots,M-1,
\end{align}
where $\widetilde{U}^{m-1} = (S^m)^\dagger W^m$. The contributions to the gradient from the final window follow from
\begin{align}\label{eq_dT-dW-m-1}
    \frac{\p P_M}{\p W^{M-1}_{r,pq}} = 
    \frac{2}{n^2} \mbox{Re} \left( \left\langle \eb_p \eb_q^\dagger, nW^{M-1} - \gamma_M (S^M)^\dagger {\cal V} \right\rangle_F \right)
    = \frac{2}{n^2} \mbox{Re} \left\{ n W^{M-1} - \gamma_M \widetilde{U}^{M-1} \right\}_{pq},
\end{align}
and, similarly,
\begin{align}\label{eq_dT-dW-m-1-im}
    \frac{\p P_M}{\p W^{M-1}_{i,pq}} = 
    \frac{2}{n^2} \mbox{Im} \left\{ n W^{M-1} - \gamma_M \widetilde{U}^{M-1} \right\}_{pq}. 
\end{align}

To evaluate the above formulae, we need to calculate $\mu\widetilde{U}^{m-1} = \mu(S^m)^\dagger W^m$ for $m=2,\ldots,M-1$ and $\gamma_M\widetilde{U}^{M-1} = \gamma_M(S^M)^\dagger {\cal V}$. Both expressions correspond to back-propagation of a state from time $t_m$ to $t_{m-1}$. Back-propagating the solution of Schr\"odinger's equation is equivalent to solving the adjoint state equation \eqref{eq_lambda-de} subject to terminal conditions. In particular, the terminal conditions from \eqref{eq_terminal-app} result in
\begin{align}
    \Lambda^m(t_{m-1}) = (S^m)^\dagger \Xi_m = \begin{cases}
        \frac{\mu}{2} (S^m)^\dagger\left(W^m - U^m(t_m) \right),\quad & m=1,\ldots,M-1,\\
        \frac{1}{n^2} (S^M)^\dagger\left(\gamma_M{\cal V} - n U^M(t_M) \right),& m=M.
    \end{cases}
\end{align}
Here, $(S^m)^\dagger U^m(t_m) = W^{m-1}$ because $U^m(t_m) = S^m W^{m-1}$. Therefore,
\begin{align}
    \frac{\mu}{2} \widetilde{U}^{m-1} = \frac{\mu}{2} (S^m)^\dagger W^m &= \Lambda^m(t_{m-1}) + \frac{\mu}{2} W^{m-1},\quad m=1,\ldots,M-1,\label{eq_back-prop-1}\\
    \frac{\gamma_M}{n^2} \widetilde{U}^{M-1} = \frac{\gamma_M}{n^2}(S^M)^\dagger{\cal V} &= \Lambda^M(t_{M-1}) + \frac{1}{n} W^{M-1}.\label{eq_back-prop-2}
\end{align}
After substituting \eqref{eq_back-prop-1} into \eqref{eq_dT-dW-j-1} and \eqref{eq_back-prop-2} into \eqref{eq_dT-dW-m-1}-\eqref{eq_dT-dW-m-1-im}, we arrive at the unified formulae
\begin{align}
    \frac{\p P_m}{\p W^{m-1}_{x,pq}} &= 
    -2 \left\{  \Lambda^m_x(t_{m-1}) \right\}_{pq},\quad m=2,\ldots,M,\quad x=\{r,i\}.\label{eq_dT-dW-j}
\end{align} 
Since the gradient with respect to $W^m$ gets a contributions from the terms $P_m$ and $P_{m+1}$, combining \eqref{eq_dT-dW-j-1} and \eqref{eq_dT-dW-j} yields the final expression of the gradient in \eqref{eq_dP-dW}.

\bibliographystyle{plain}
\bibliography{mybib,quantum}

\begin{thebibliography}{10}

\bibitem{PetGar-22}
N.~Anders~Petersson and Fortino Garcia.
\newblock Optimal control of closed quantum systems via {B}-splines with
  carrier waves.
\newblock {\em SIAM Journal on Scientific Computing}, 44(6):A3592--A3616, 2022.

\bibitem{petsc-web-page}
Satish Balay, Shrirang Abhyankar, Mark~F. Adams, Jed Brown, Peter Brune, Kris
  Buschelman, Lisandro Dalcin, Alp Dener, Victor Eijkhout, William~D. Gropp,
  Dinesh Kaushik, Matthew~G. Knepley, Dave~A. May, Lois~Curfman McInnes,
  Richard~Tran Mills, Todd Munson, Karl Rupp, Patrick Sanan, Barry~F. Smith,
  Stefano Zampini, Hong Zhang, and Hong Zhang.
\newblock {PETS}c {W}eb page, 2020.
\newblock https://www.mcs.anl.gov/petsc.

\bibitem{borzi2011computational}
Alfio Borz{\`\i} and Volker Schulz.
\newblock {\em Computational optimization of systems governed by partial
  differential equations}.
\newblock SIAM, 2011.

\bibitem{Caneva-2011}
T.~Caneva, T.~Calarco, and S.~Montangero.
\newblock Chopped random-basis quantum optimization.
\newblock {\em Physical Review A}, 84(2), Aug 2011.

\bibitem{ChuFre-22}
Seung~Whan Chung and Jonathan~B. Freund.
\newblock An optimization method for chaotic turbulent flow.
\newblock {\em Journal of Computational Physics}, 457:111077, 2022.

\bibitem{CombarizaEtAl-91}
J.~E. Combariza, B.~Just, J.~Manz, and G.~K. Paramonov.
\newblock Isomerizations controlled by ultrashort infrared laser pulses: model
  simulations for the inversion of ligands (h) in the double-well potential of
  an organometallic compound, [(c5h5)(co)2feph2].
\newblock {\em The Journal of Physical Chemistry}, 95(25):10351--10359, 1991.

\bibitem{Debnath-16}
S.~Debnath, N.~Linke, C.~Figgatt, K.~A. Landsman, K.~Wright, and C.~Monroe.
\newblock Demonstration of a small programmable quantum computer with atomic
  qubits.
\newblock {\em Nature}, 536:63–66, 2016.

\bibitem{Emsley-1989}
L.~Emsley and G.~Bodenhausen.
\newblock Gaussian pulse cascades: New analytical functions for rectangular
  selective inversion and in-phase excitation in {NMR}.
\newblock {\em Chem. Phys.}, 165(6):469--476, 1989.

\bibitem{Ewing-1990}
B.~Ewing, S.~J. Glaser, and G.~P. Drobny.
\newblock Development and optimization of shaped {NMR} pulses for the study of
  coupled spin systems.
\newblock {\em Chem. Phys.}, 147:121--129, 1990.

\bibitem{Fang-22}
Liang Fang, Stefan Vandewalle, and Johan Meyers.
\newblock A parallel-in-time multiple shooting algorithm for large-scale
  pde-constrained optimal control problems.
\newblock {\em Journal of Computational Physics}, 452:110926, 2022.

\bibitem{glaser2015training}
Steffen~J Glaser, Ugo Boscain, Tommaso Calarco, Christiane~P Koch, Walter
  K{\"o}ckenberger, Ronnie Kosloff, Ilya Kuprov, Burkhard Luy, Sophie Schirmer,
  Thomas Schulte-Herbr{\"u}ggen, et~al.
\newblock Training schr{\"o}dinger’s cat: quantum optimal control.
\newblock {\em The European Physical Journal D}, 69(12):1--24, 2015.

\bibitem{goerz2019krotov}
Michael~H Goerz, Daniel Basilewitsch, Fernando Gago-Encinas, Matthias~G Krauss,
  Karl~P Horn, Daniel~M Reich, and Christiane~P Koch.
\newblock Krotov: A python implementation of krotov's method for quantum
  optimal control.
\newblock {\em SciPost physics}, 7, 2019.

\bibitem{quandaryGithub}
Stefanie G{\"u}nther and N.~Anders Petersson.
\newblock Quandary: Optimal control for open quantum systems.
\newblock Github, 2021.

\bibitem{gunther2021quandary}
Stefanie G{\"u}nther, N~Anders Petersson, and Jonathan~L DuBois.
\newblock Quandary: an open-source c++ package for high-performance optimal
  control of open quantum systems.
\newblock In {\em 2021 IEEE/ACM Second International Workshop on Quantum
  Computing Software (QCS)}, pages 88--98. IEEE, 2021.

\bibitem{JanMey-24}
N.~Janssens and J.~Meyers.
\newblock Parallel-in-time multiple shooting for optimal control problems
  governed by the navier–stokes equations.
\newblock {\em Computer Physics Communications}, 296:109019, 2024.

\bibitem{Khaneja-2005}
N.~Khaneja, T.~Reiss, C.~Kehlet, T.~Schulte-Herbruggen, and S.~Glaser.
\newblock Optimal control of coupled spin dynamics: design of {NMR} pulse
  sequences by gradient ascent algorithms.
\newblock {\em J. Magnetic Resonance}, 172:296--305, 2005.

\bibitem{koch2022quantum}
Christiane~P Koch, Ugo Boscain, Tommaso Calarco, Gunther Dirr, Stefan Filipp,
  Steffen~J Glaser, Ronnie Kosloff, Simone Montangero, Thomas
  Schulte-Herbr{\"u}ggen, Dominique Sugny, et~al.
\newblock Quantum optimal control in quantum technologies. strategic report on
  current status, visions and goals for research in europe.
\newblock {\em EPJ Quantum Technology}, 9(1):19, 2022.

\bibitem{Leung-2017}
N.~Leung, M.~Abdelhafez, Jens Koch, and D.~Schuster.
\newblock Speedup for quantum optimal control from automatic differentiation
  based on graphics processing units.
\newblock {\em Phys. Rev. A}, 95:0432318, 2017.

\bibitem{Lucarelli-2018}
Dennis Lucarelli.
\newblock Quantum optimal control via gradient ascent in function space and the
  time-bandwidth quantum speed limit.
\newblock {\em Physical Review A}, 97(6), Jun 2018.

\bibitem{Morzhin2018KrotovMF}
Oleg~V. Morzhin and Alexander~N. Pechen.
\newblock Krotov method for optimal control of closed quantum systems.
\newblock {\em Russian Mathematical Surveys}, 74:851 -- 908, 2018.

\bibitem{nocedal2006numerical}
Jorge Nocedal and Stephen Wright.
\newblock {\em Numerical optimization}.
\newblock Springer Science \& Business Media, 2006.

\bibitem{Brumer-92}
M~Shapiro P~Brumer.
\newblock Laser control of molecular processes.
\newblock {\em Annual review of physical chemistry}, 43(1):257--282, 1992.

\bibitem{PucBaiBeg-23}
P.~Puchaud, F.~Bailly, and M.~Begon.
\newblock Direct multiple shooting and direct collocation perform similarly in
  biomechanical predictive simulations.
\newblock {\em Computer Methods in Applied Mechanics and Engineering},
  414:116162, 2023.

\bibitem{ShiRab-92}
Shenghua Shi and Herschel Rabitz.
\newblock Optimal control of selectivity of unimolecular reactions via an
  excited electronic state with designed lasers.
\newblock {\em The Journal of Chemical Physics}, 97(1):276--287, 1992.

\bibitem{smith2022programming}
Kaitlin~N Smith, Gokul~Subramanian Ravi, Thomas Alexander, Nicholas~T Bronn,
  Andr{\'e}~RR Carvalho, Alba Cervera-Lierta, Frederic~T Chong, Jerry~M Chow,
  Michael Cubeddu, Akel Hashim, et~al.
\newblock Programming physical quantum systems with pulse-level control.
\newblock {\em Frontiers in physics}, 10:900099, 2022.

\bibitem{mpibook}
Marc Snir, Steve Otto, Steven Huss-Lederman, David Walker, and Dongarra Jack.
\newblock {\em {MPI}: {T}he {C}omplete {R}eference}.
\newblock The MIT Press, 1998.

\bibitem{spiteri2018quantum}
Raymond~J Spiteri, Marina Schmidt, Joydip Ghosh, Ehsan Zahedinejad, and Barry~C
  Sanders.
\newblock Quantum control for high-fidelity multi-qubit gates.
\newblock {\em New Journal of Physics}, 20(11):113009, 2018.

\bibitem{Trowbridge-23}
A.~Trowbridge, A.~Bhardwaj, K.~He, D.~I. Schuster, and Z.~Manchester.
\newblock Direct collocation for quantum optimal control.
\newblock In {\em 2023 IEEE International Conference on Quantum Computing and
  Engineering (QCE)}, pages 1278--1285, Los Alamitos, CA, USA, sep 2023. IEEE
  Computer Society.

\bibitem{Unser97}
M.~Unser.
\newblock Ten good reasons for using spline wavelets.
\newblock {\em Wavelet Applications in Signal and Image Processing V}, 3169, 09
  2002.

\bibitem{Wachter2006}
A.~W{\"a}chter and L.~T. Biegler.
\newblock On the implementation of an interior-point filter line-search
  algorithm for large-scale nonlinear programming.
\newblock {\em Mathematical Programming}, 106(1):25--57, Mar 2006.

\bibitem{Zax-1988}
D.~B. Zax, G.~Goelman, and S.~Vega.
\newblock Amplitude-modulated composite pulses.
\newblock {\em J. Magn. Reson.}, 80(2):375--382, 1988.

\end{thebibliography}

\end{document}